\documentclass[prd,aps,nofootinbib,
showkeys]
{revtex4}
\usepackage{graphicx,epsf,amsfonts,amssymb,amsbsy}
\newcommand{\mat}{\left ( \begin{array}}
\newcommand{\emat}{\end{array} \right )}
\newcommand{\vect}{\left ( \begin{array}{c}}
\newcommand{\evect}{\end{array} \right )}

\begin{document}

\title{Weak dual symmetries of two color QCD phase diagram} 

\def\addressa{address 1}
\def\addressb{address 2}

\author{\firstname{K. G.}~\surname{Klimenko}}
\affiliation{State Research Center
of Russian Federation -- Institute for High Energy Physics,
NRC "Kurchatov Institute", 142281 Protvino, Moscow Region, Russia}
\author{\firstname{R. N.}~\surname{Zhokhov}}
\email[E-mail: ]{zhokhovr@gmail.com}
\affiliation{Pushkov Institute of Terrestrial Magnetism, Ionosphere and Radiowave Propagation (IZMIRAN),
108840 Troitsk, Moscow, Russia}

\received{xx.xx.2025}
\revised{xx.xx.2025}
\accepted{xx.xx.2025}

\begin{abstract}
The phase diagram of two color QCD under influence of baryon density ($\mu_B$), isospin ($\mu_I$), chiral ($\mu_5$ and $\nu_5$) imbalances is investigated within the framework NJL model approach. This letter establishes the existence of weak dualities in two color quark matter—symmetries revealed when full thermodynamic potential is projected or restricted to specific condensation channels.
These dualities provide a unified understanding of two interesting phenomena in the phase diagram: universal catalysis and chameleon effect of chiral chemical potential.
\end{abstract}

\pacs{Suggested PACS}\par
\keywords{Suggested keywords   \\[5pt]}

\maketitle


\section{Introduction}
The phase structure of Quantum Chromodynamics (QCD) at finite temperature $T$ and baryon chemical potential $\mu_B$ represents a cornerstone of modern nuclear and particle physics, delineating the landscape of strong-interaction matter under extreme conditions \cite{Kogut:2004zz, Fukushima:2010bq}. At high temperatures and low baryon density, lattice QCD calculations have conclusively demonstrated a crossover transition from a hadronic phase, where chiral symmetry is spontaneously broken and color is confined, to a deconfined quark gluon matter phase where chiral symmetry is approximately restored \cite{Aoki:2009sc, Borsanyi:2020fev}. This high-$T$, low-$\mu_B$ region is directly probed in high energy heavy-ion collision experiments.
In stark contrast, the high baryon density and low temperature regime remains largely terra incognita, presenting a formidable challenge to first principle methods such as lattice QCD, where the so-called fermion sign problem renders standard lattice QCD simulations intractable at finite $\mu_B$ \cite{Aarts:2015tyj}. This region of the phase diagram is theorized to host a rich array of exotic phases, including color superconducting states featuring diquark condensation \cite{Alford:2007xm}, a critical point marking the end of the crossover line, etc. Consequently, our understanding is driven by a combination of effective models, perturbative calculations at asymptotically high densities, and the exploration of closely related theories and models.

This theoretical impediment necessitates the investigation of models and related gauge theories that serve as analogues, offering insights into the universal features of dense quark matter. So here comes two color Quantum Chromodynamics (QC$_2$D)—the non-Abelian gauge theory with $SU(2)$ color symmetry group and $N_f$ fermion flavors (here we assume $N_f=2$). Despite its distinct gauge group, two color QCD exhibits numerous qualitative features common with real QCD, including confinement, chiral symmetry breaking, and the emergence of a superfluid phase at high baryon density with diquark condensation akin to color superconductivity. Also its advantage lies in its accessibility to first principle lattice simulations, as it is exempt from the sign problem at non-zero baryon chemical potential \cite{Bragutasymmetry}. This allows for ab initio computation of the equilibrium thermodynamic properties and the determination of its phase diagram at non-zero baryon density with controlled systematic errors. Due to all these properties it provides a vital theoretical laboratory for studying the phase structure of strong interactions at finite density.

The standard mapping of the QCD phase diagram via the temperature $T$ and baryon chemical potential $\mu_B$ represents only a part of a richer multidimensional parameter space. In recent years, this picture has been expanded by considering more types of chemical potentials. This includes the isospin chemical potential $\mu_I$ \cite{Ayala:2024sqm, Lopes:2025rvn, Ayala:2023mms, Basta:2025svw, Brandt:2024dle, Andersen:2023ivj, AndersenBrauner, AndersenKyllingstad, Zhuang, AndersenKneschke,Basta:2025svw,Andersen:2025ezj,Ayala:2024sqm,Lopes:2025rvn, Khunjua:2020hbd, Khunjua:2024qha}, as well as the chiral chemical potentials $\mu_5$ and $\mu_{I5}$ \cite{Azeredo:2024sqc, braguta,Braguta:2015zta, bragutakotov, bragutakatsnelson, RuggieriGatto, Huang, Huang1, RuggieriPeng, andrianov, espriu, farias, Zhuang1, Farias:2016let}, which introduce imbalance between the number of left- and right-handed fermions.

Recently it has been shown that there exist dual correspondence between chiral symmetry breaking and charged pion condensation phenomena and quark matter with isospin $\mu_I$ and chiral $\mu_{I5}$ chemical potentials \cite{kkz18-2, kkz18, CSC}. Then it was found that there exist even more dualities in the two color case \cite{Khunjua:2020xws, Khunjua:2021oxf}. Then dualities were shown from first principles as in two color QCD as well as in three color one \cite{Khunjua:2024kdc, Klimenko:2024qzq}

\section{Two color NJL model}

The Lagrangian of effective NJL model that describes the basic low-energy properties of dense ($\mu_B$) two-color quark matter with isospin ($\mu_I$) and chiral asymmetries ($\mu_5$ and $\mu_{I5}$) are
\begin{eqnarray}
L&=&\bar q \Big [i\hat\partial-m_0\Big ]q +\bar q {\cal M}\gamma^0 q+H\Big [(\bar qq)^2+(\bar qi\gamma^5\vec\tau q)^2+
\big (\bar qi\gamma^5\sigma_2\tau_2q^c\big )\big (\overline{q^c}i\gamma^5\sigma_2\tau_2 
q\big )\Big],
\label{njl}
\end{eqnarray}
where the term ${\cal M}$ is defined as 
${\cal M}= \mu+\nu\tau_3+\nu_5\gamma^5\tau_3+\mu_5\gamma^5,$ 
here chemical potentials are 

$\mu=\mu_B/2\;$ - quark and baryon chemical potential,
  
$\nu=\mu_I/2\;$ - isospin chemical potential,

$\nu_5=\mu_{I5}/2\;$ - chiral isospin chemical potential,

$\mu_5\;$ - chiral chemical potential.

It is rather reasonable to start from a semibosonized version of the Lagrangian (see details in \cite{Khunjua:2020xws, Khunjua:2021oxf})
with auxiliary bosonic fields $\sigma (x)$, $\vec\pi  =(\pi_1 (x),\pi_2 (x),\pi_3 (x))$, $\Delta (x)$ and $\Delta^* (x)$, that have the following the equations of motion
\begin{eqnarray}
\sigma (x)=-2H(\bar qq),&~~~~~&\Delta (x)= -2H\Big [\overline{q^c}i\gamma^5\sigma_2\tau_2 q\Big ]=-2H\Big [q^TCi\gamma^5\sigma_2\tau_2 q\Big ],\nonumber\\
~\vec\pi(x)=-2H(\bar qi\gamma^5\vec\tau q),&~~~~~&
\Delta^*(x)=-2H\Big [\bar qi\gamma^5\sigma_2\tau_2q^c\Big ]=-2H\Big [\bar qi\gamma^5\sigma_2\tau_2C\bar q^T\Big ].\nonumber
\end{eqnarray}

Due to a symmetry of the model, the number of condensates that characterize the ground state of a system may be reduced to only three, $M,\pi_1$ and $\Delta$, condensates.
There are several phases in the model,
chiral symmetry breaking (CSB), pion condensation (PC), baryon superfluid (BSF) and symmetric phase,

\begin{center}
\begin{itemize}
\item 
$\sigma\neq0,~\vec\pi_1=0,~\Delta=0$ ~~~~ ---  CSB phase

\item 
$\sigma=0,~\pi_1\neq0,~\Delta=0$ ~~~ --- charged PC phase

\item 
$\sigma=0,~\pi_1=0,~\Delta\neq0$ ~~~ --- BSF phase

\item 
$\sigma=0,~\pi_1=0,~\Delta=0$ ~~~ --- symmetric SYM phase.
\end{itemize}
\end{center}

\section{Dual properties}
One could show that the thermodynamic potential (TDP) and hence the whole phase diagram is invariant with respect to the so-called dual transformations ${\cal D}_1$, ${\cal D}_2$ and ${\cal D}_3$
\begin{eqnarray}
{\cal D}_1: ~\mu\longleftrightarrow\nu,~~~\pi_1\longleftrightarrow |\Delta|,~~~~~~
{\cal D}_2: ~\mu\longleftrightarrow\nu_5,~~M\longleftrightarrow |\Delta|,~~~~~~{\cal D}_3:
~\nu\longleftrightarrow\nu_5,~~M\longleftrightarrow \pi_1.\nonumber
\end{eqnarray}
One can see that chemical potentials $\mu$, $\mu_{I}$ and $\mu_{I5}$ are connected by dual transformations.

\section{Weak dualities}
Now let us recall that in the three color case it was demonstrated that there exist so called weak dualities \cite{weakdual}. These dualities refer to symmetries in the phase diagram that become apparent when the TDP is projected onto the axes corresponding to individual condensates. So in order to investigate weak dualities one need to consider the projections of the TDP on various axes.

In effective NJL model (\ref{njl}), when the diquark condensate is absent ($\Delta = 0$), the TDP depends on the quantity $P_+(\eta)P_-(\eta)$ (we denote $\eta=p_0+\mu$), where
\begin{eqnarray}
P_\pm(\eta)&\equiv&\pi_1^4-2\pi_1^2\big [(p_0+\mu)^2-(|\vec p|\pm\mu_5)^2-M^2+\nu_5^2-\nu^2\big ]\nonumber\\
&+&\big [M^2+(|\vec p|\pm\mu_5+\nu_5)^2-(p_0+\mu\pm\nu)^2\big ]\big [M^2+(|\vec p|\pm\mu_5-\nu_5)^2-
(p_0+\mu\mp\nu)^2\big ]. \label{17}
\end{eqnarray}

Additionally restricting to the case of a vanishing pion condensate ($\pi = 0$), one can obtain that the TDP depends on
\begin{eqnarray}
&&P_M(\mu,\, \nu,\, \nu_5,\, \mu_5)=\ln P_+(\eta)P_-(\eta)\Big |_{\Delta=0,\, \pi=0}=\ln\Big\{\big [M^2+(|\vec p|+\mu_5+\nu_5)^2-(p_0+\mu+\nu)^2\big ]\times\nonumber\\
&&
\times\big [M^2+(|\vec p|+\mu_5-\nu_5)^2-
(p_0+\mu-\nu)^2\big ]
\big [M^2+(|\vec p|-\mu_5+\nu_5)^2-(p_0+\mu-\nu)^2\big ]\times\nonumber\\
&&\times\big [M^2+(|\vec p|-\mu_5-\nu_5)^2-(p_0+\mu+\nu)^2\big ]\Big\}.
\label{18}
\end{eqnarray}

One can see that the quantity $P_M$ exhibits the following property $P_M(\mu,\, \nu,\, \nu_5,\, \mu_5)=P_M(\nu,\,\mu ,\, \nu_5,\, \mu_5)$. One can also notice that for the invariance it can be shown that one can change the sign of $p_0$ (or one could argue just by dual property ${\cal D}_1$). 
So projection of the TDP on $M$ axis that depends on $P_M$ should possess this property, i. e it is invariant with respect to the transformation
\begin{eqnarray}
\mu\longleftrightarrow\nu.
\end{eqnarray}

Moreover, it is possible to note that the quantity $P_M$ satisfies: $P_M(\mu,\, \nu,\, \nu_5,\, \mu_5)=P_M(\mu,\, \nu,\,  \mu_5,\,\nu_5)$. 
And hence the projection of the TDP on $M$ axis 
$\Omega(M |\, \mu,\, \nu,\,  \nu_5,\, \mu_5)$ is invariant under transformation
\begin{eqnarray}
{\cal D_M}:~~~\mu_5\longleftrightarrow\nu_{5}.
 \label{DBM}
\end{eqnarray}

The invariance \ref{DBM} is called weak dual symmetry of the phase diagram. It represents a symmetry that is not present in the full TDP but emerges in the case when it is enough to consider the projection onto a specific order parameter. 
We will discuss the implications and usage in the next section.

We now consider the phase structure under the condition of restored chiral symmetry (chiral condensate is zero, $M=0$) and in the absence of diquark condensation ($\Delta=0$). This scenario is physically relevant, for example, for systems with high isospin density, where pion condensation is expected to be the dominant phenomenon. In this case, the TDP depends on the quantity
\begin{eqnarray}
&&P_\pi(\mu,\, \nu,\, \nu_5,\, \mu_5)=\big [\pi_1^2+(|\vec p|+\mu_5+\nu)^2-(\eta+\nu_5)^2\big ]
\big [\pi_1^2+(|\vec p|+\mu_5-\nu)^2-
(\eta-\nu_5)^2\big ]\nonumber\\
&&
\big [\pi_1^2+(|\vec p|-\mu_5+\nu)^2-(\eta-\nu_5)^2\big ]\big [\pi_1^2+(|\vec p|-\mu_5-\nu)^2-(\eta+\nu_5)^2\big ].
\label{20}
\end{eqnarray}


A direct analysis of this expression reveals a symmetry under the interchange of the chemical potentials $\nu$ and $\mu_5$.
Consequently, the projection of the TDP onto the pion condensate axis, $\Omega(\pi | \mu, \nu, \nu_5, \mu_5)$, must inherit this invariance.
So in situations if chiral symmetry breaking phenomenon and color superconductivity phenomena can be omitted and only pion condensation matters the TDP satisfies the dual symmetry
\begin{eqnarray}
{\cal D}_{\pi}:~~~~\mu_5\longleftrightarrow\nu.
 \label{DBpi}
\end{eqnarray}
This establishes a weak dual symmetry in the sector dominated by pion condensation. 

Finally, we consider quark matter with diquark condensation, i. e. under the condition of restored chiral symmetry ($M=0$) and in the absence of pion condensation ($\pi_1=0$). In this case, the TDP of the model would depend on the following quantity
\begin{eqnarray}
&&P_\Delta(\mu,\, \nu,\, \nu_5,\, \mu_5)=\ln P_+^{{\small \cal \Delta}}(\eta)P_-^{\small \cal \Delta}(\eta)\Big |_{M=0,\, \pi=0}=\ln\Big\{\big [\Delta^2+(|\vec p|+\mu_5+\mu)^2-(p_0+\nu_5+\nu)^2\big ]\times\nonumber\\
&&
\times\big [\Delta^2+(|\vec p|+\mu_5-\mu)^2-
(p_0+\mu_5-\nu)^2\big ]
\big [\Delta^2+(|\vec p|-\mu_5+\mu)^2-(p_0+\mu-\nu)^2\big ]\times\nonumber\\
&&\times\big [\Delta^2+(|\vec p|-\mu_5-\mu)^2-(p_0+\mu+\nu)^2\big ]\Big\}.
\label{18}
\end{eqnarray}

\begin{figure}
\includegraphics[width=0.75\textwidth]{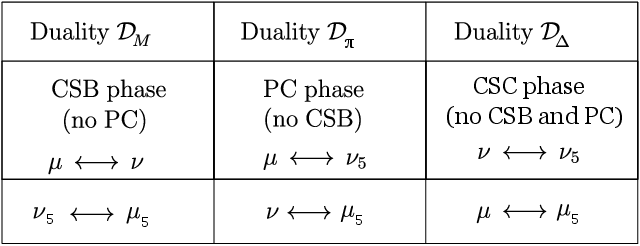}
\caption{\textbf{Schematic structure of weak dualities}}
\end{figure}

The structure of the quantity $P_{\Delta}(\mu, \nu, \nu_5, \mu_5)$ governing the diquark condensation phenomenon in dense quark matter reveals several key dual symmetries.

It is possible to show (one either need to use invariance under $p_0\to-p_0$ or just employ one of the main or fundamental dualities, i. e. ${\cal D}_3$) that the quantity $P_{\Delta}$ respect the relation with exchange of chiral chemical potentials 
$
P_{\Delta}(\mu,\, \nu,\, \nu_5,\, \mu_5)=P_M(\nu,\,\mu ,\, \nu_5,\, \mu_5)
$
and possesses the following dual symmetry
\begin{eqnarray}
\nu\longleftrightarrow\nu_5
\end{eqnarray}

Furthermore, one can note that another intrinsic symmetry of $P_{\Delta}$ can be identified. A direct analysis demonstrates that it is invariant under the interchange of baryon and chiral (axial)  chemical potentials.
This implies that the projection of the TDP onto the diquark axis, $\Omega(\Delta | \mu, \nu, \nu_5, \mu_5)$, possesses the weak duality 
\begin{eqnarray}
{\cal D}_{ \Delta}:~~~~\mu\longleftrightarrow\mu_5.
\label{DBDelta}
\end{eqnarray}
This establishes weak duality for the diquark condensation phenomena of quark matter.

For convenience the schematic scheme of weak dual symmetries of 
the phase diagram of two color quark matter is depicted in Fig. 1


\section{Phase diagram and weak dual symmetries}

There have been inferred several unique inherent property of chiral chemical potential $\mu_5$ in terms of its effect on the phase structure of quark matter. Let us discuss it and find the connections with weak dual symmetries and show that these interesting features are just the consequences of these dual symmetries.

\subsection{Universal catalysis}

Let us first discuss the phase structure of quark matter with non-zero chiral imbalances.
A rather well established result in this context is the enhancement of chiral symmetry breaking by non-zero chiral chemical potentials. This effect was called chiral catalysis of chiral symmetry breaking and was predicted for the chiral chemical potential $\mu_5$ within effective models \cite{Farias:2016let} (in some papers the opposite effect of anti-catalysis was also predicted, it was not clear first) and subsequently observed in lattice simulations \cite{Braguta:2015zta}. An analogous enhancement driven by different chiral imbalance, namely the chiral (axial) isospin chemical potential $\nu_5$, was later predicted as well \cite{Khunjua:2018dbm, Khunjua:2018jmn}.
For example, in terms of effective NJL model one can see that chiral condensate rises then chiral imbalance $\nu_5$ is increased. In a similar fashion one can show that chiral condensate increases by the same rate if one considers non-zero value of chiral chemical potential $\mu_5$. 
In Fig. 2 one can see the behaviour of chiral condensate as a function of chiral chemical potentials $\mu_5$ and $\nu_5$ (the behavior of condensate shown in Fig. 2 (c) is typical for any fixed value of $\mu_5$ and $\nu_5$ including zero value, $100$ MeV is shown just for illustration purposes). The observed effect is not a mere coincidence but this picture is direct consequence of weak duality property of two color quark matter (\ref{DBM}).

\begin{figure}
\includegraphics[width=1.0\textwidth]{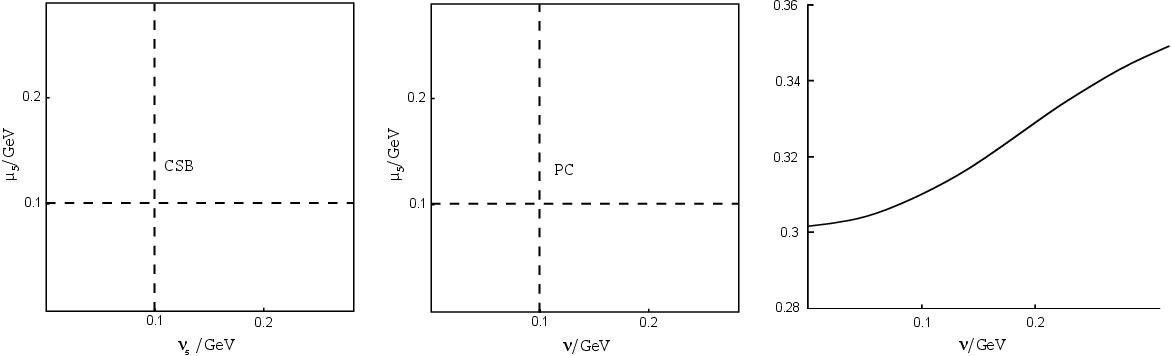}
\caption{\textbf{(a) ($\nu_5$, $\mu_5$)-phase diagram at $\mu=0$ and $\nu=0$ (b) ($\nu$, $\mu_5$)-phase diagram at $\mu=0$ and $\nu_5=0$ (c) Order parameter $\pi_1$ as a function of $\nu$ at $\mu_5=100$ MeV, $\mu=0$ and $\nu_5=0$ and as a function of $\mu_5$ at $\nu=100$ MeV, $\mu=0$ and $\nu_5=0$}}
\end{figure}


We now turn to the phase structure of quark matter with a non-zero isospin imbalance, characterized by an isospin chemical potential $\nu$ (or equivalently, $\mu_I = 2\nu$). Also further let us assume that we work in the chiral limit, i. e. zero value of current quark mass $m_0=0$, implying zero value of pion mass $m_{\pi}=0$. The system is known to undergo a phase transition to a pion condensation phase at a critical value of the isospin chemical potential $\nu=m_\pi/2$. Beyond this threshold, the pion condensate increases with $\nu$, demonstrating that isospin imbalance catalyzes pion condensation a well established result in both effective models \cite{AndersenKyllingstad} and lattice QCD \cite{Brandt:2024dle, Brandt:2023kev} (the latter for $m_\pi \neq 0$).
A more intricate phenomenon emerges when both isospin and chiral imbalances are present. In the chiral limit introducing even an infinitesimally small isospin chemical potential $\nu$ induces pion condensation. Furthermore, increasing $\mu_5$ enhances the pion condensate, indicating that chiral imbalance also acts as a catalyst for pion condensation under these conditions.
This catalytic effect is not an independent phenomenon but a direct and inevitable consequence of the weak dual symmetry (\ref{DBpi}). This duality, which enforces an equivalence between the roles of $\nu$ and $\mu_5$ in the pion condensation sector of the thermodynamic potential, mandates that the system response to both chemical potentials must be identical.
This behavior is illustrated in Fig. 2 (c), which shows the pion condensate as a function of $\nu$ for a fixed $\mu_5 = 0.1$ GeV, and as a function of $\mu_5$ for a fixed $\nu = 0.1$ GeV. 
These curves represent specific one dimensional sections (as indicated by the dashed lines in Fig. 2 (b)) of the broader $(\nu, \mu_5)$-phase diagram. The choice of $100$ MeV is merely representative, the same qualitative behavior holds for arbitrary values of these chemical potentials. The identical functional form of these dependencies is not accidental, but 
rigorously required by the weak duality (\ref{DBpi}).


\begin{figure}
\includegraphics[width=1.0\textwidth]{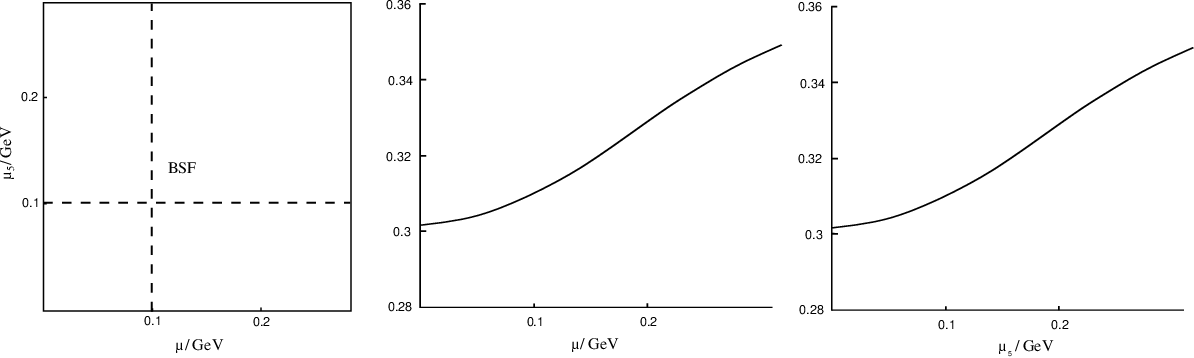}
\caption{\textbf{(a) ($\mu$, $\mu_5$)-phase diagram at $\nu=0$ and $\nu_5=0$ (b) Order parameter $\Delta$ as a function of $\mu$ at $\mu_5=100$ MeV, $\nu=0$ and $\nu_5=0$ (c) Order parameter $\Delta$ as a function of $\mu_5$ at $\mu=100$ MeV, $\nu=0$ and $\nu_5=0$}}
\end{figure}

Now let us discuss baryon superfluid (BSF) phenomenon in two color QCD, which occurs at non-zero baryon density. 
In the chiral limit, the BSF phase becomes dominant at any infinitesimally small but non-zero value of baryon chemical potential $\mu$. As illustrated in Fig. 3 (b), the diquark condensate
 increases steadily with further increase of $\mu$.
A key finding is that a chiral imbalance, introduced via the chiral chemical potential $\mu_5$, 
further enhances the diquark condensate, meaning strengthening the BSF phase and increasing the associated energy gap. This catalytic effect arises directly from the weak duality (\ref{DBDelta}). Crucially, the enhancement of the diquark condensate with increasing $\mu_5$ is of precisely identical functional dependence as with increase of $\mu$. 
This is clearly demonstrated in Fig. 3 (b) and (c), at $\mu = 100$ MeV and $\mu_5 = 100$ MeV, the BSF phase is first generated by $\mu$ and is then further catalyzed in the same fashion as by $\mu$ as well as by $\mu_5$. 

The discovery of the full set of weak dual symmetries implies that $\mu_5$ can catalyze any condensation phenomenon in the system. Each chemical potential first generates its associated phenomenon (e.g., $\mu$ generates BSF, $\nu$ generates PC), and increasing its value enhances its respective condensate. The weak dualities reveal that $\mu_5$ is functionally equivalent to each of these chemical potentials in different sectors of the model. Therefore, an increase in $\mu_5$ will catalyze the chiral condensate (due to (\ref{DBM})), the pion condensate (due to (\ref{DBpi})), and the diquark condensate (due to (\ref{DBDelta})) with the same efficiency as their respective conjugate potentials.

\subsection{Chameleon like property of chiral imbalance}

\begin{figure}
\includegraphics[width=1.0\textwidth]{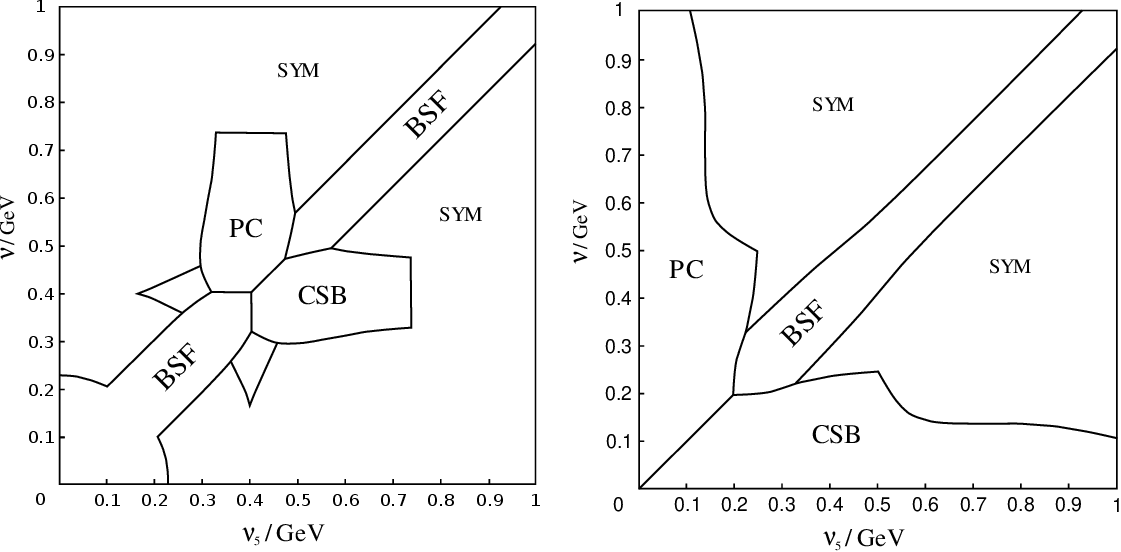}
\caption{\textbf{($\nu_5$, $\nu$)-phase diagram (a) at $\mu=400$ MeV and $\mu_5=0$; (b) ($\nu_5$, $\nu$)-phase diagram at $\mu_5=400$ MeV and $\mu=0$.}}
\end{figure}

As discussed in the previous subsection, in quark matter with non-zero baryon density and non-zero chiral imbalance, where the BSF phase is realized, the system responds identically to further increases of either chiral chemical potential $\mu_5$ or baryon one $\mu$, the BSF phase is catalyzed in exactly the same fashion. If a small baryon chemical potential is present (see Fig. 3 (a)), increasing $\mu_5$ from zero to large values produces the same enhancement in the diquark condensate as achieved by increasing $\mu$ to equally large values (and small $\mu_5$).
In the chiral limit, $\mu$ can be arbitrarily small (just sufficient to cause BSF phase), the system cannot distinguish between the presence of a large baryon density $n_B$ or a large chiral imbalance $n_5$. This is the same picture observed with the chiral chemical potentials $\mu_5$ and $\nu_5$: starting from a vacuum (with all chemical potentials zero) where the system is in the chirally broken CSB phase, introducing and increasing either $\nu_5$ or $\mu_5$ enhances the chiral condensate to the same degree (see Fig. 2 (a)).
One can conclude that in this scenario the quark matter does not feel the difference between baryon chemical potential and chiral chemical potential $\mu_5$ and that $\mu_5$ can play the role of $\mu$ in this conditions. In a similar way that chiral chemical potentials $\nu_5$ or $\mu_5$ if acts alone exhibit the same effect on the phase structure of quark matter and one of them can completely play the role of the other.
Moreover, if isospin imbalance triggers pion condensation, increasing the chiral imbalance $\mu_5$ influences the system in the same way as increasing the isospin imbalance $\nu$ (see Fig. 2 (b)), despite their physically distinct origins. 
The property of $\mu_5$ that in different scenarios it can play the role of other chemical potentials can be called the chameleon property. Meaning not just leading to similar qualitative picture but it can completely substitute other chemical potentials and lead to the same values of condensates in the system.

Now let us consider scenario with several and large-value chemical potentials, which reveal the chameleon property from another perspective. 
If we consider the phase diagram in the section of isospin $\nu$ and chiral isospin $\nu_5$ imbalances in two completely different scenarios: at rather large value of $\mu$ with zero $\mu_5$ and with rather large chiral imbalance $\mu_5$ with zero baryon chemical potential $\mu=0$. In the former scenario one can see the easily anticipated behaviour that baryon chemical potential $\mu$ triggers the generation of BSF phase (see Fig. 4 (a)). It is rather expected since $\mu$ is as a rule the parameter that lead to Fermi surface instability and diquark condensation. But in the latter scenario one can see that the non-zero chiral imbalance $\mu_5$ lead to diquark condensation without any baryon chemical potential $\mu=0$ in the same fashion as baryon chemical potential does (at $\mu_5=0$). In Fig. 4 one can see that BSF phase is identical where it is realized. It is a direct implication of weak dual symmetry (\ref{DBDelta}) between baryon and chiral chemical potentials. Let us stress that other phases in the figure such as CSB and PC phases do not respect this weak duality and all the implications can be done only in the regions outside these phases. It is typical situation for weak dual symmetries that is why they are called weak since their scope is limited. But as it is seen here it is still rather large and their use is rather helpful in analyzing the phase structure and getting new unexpected and captivating results.

Furthermore, in quark matter with both baryon density and chiral imbalance $\nu_5$, chiral chemical potential $\mu_5$ can emulate the role of $\nu$ and lead to generation of pion condensation phenomenon in the system. Figure 5 (a) and (b) compares two cases: non-zero $\nu$ with $\mu_5=0$, and non-zero $\mu_5$ with $\nu=0$. 
One can see that the charged pion condensation phase is invariant under weak dual symmetry (\ref{DBpi}) that guarantee the generation of PC phase by chiral imbalance $\mu_5$ of quark matter in these conditions. The generation of chiral symmetry breaking by chiral imbalance of any form is demonstrated in Fig. 5 (c) and (d).


\section{Conclusion}
In this letter, it was found that the phase diagram of two color QCD, described by an effective NJL model, possesses weak dual symmetries. 
The weak dualities provide a unified explanation for two remarkable features of phase structure of two color quark matter.
{\it Universal catalysis}. Chiral chemical potential $\mu_5$ acts as a universal catalyzer enhancing with identical functional dependence the formation of the chiral, pion and the diquark condensates. 
{\it Chameleon property of chiral imbalance}. The chemical potential 	$\mu_5$ in different physical contexts can exactly mimic the effects of other chemical potentials.




\begin{figure}
\includegraphics[width=1.0\textwidth]{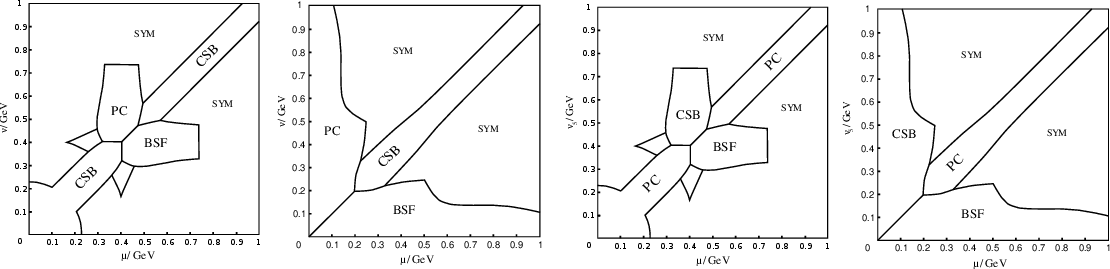}
\caption{\textbf{(a) ($\nu$, $\mu$)-phase diagram at $\nu_5=400$ MeV and $\mu_5=0$; (b) ($\nu$, $\mu$)-phase diagram at $\mu_5=400$ MeV and $\nu_5=0$; (c) ($\nu_5$, $\mu$)-phase diagram at $\nu_5=400$ MeV and $\mu_5=0$; (d)($\nu_5$, $\mu$)-phase diagram at $\mu_5=400$ MeV and $\nu_5=0$.}}
\end{figure}


\end{document}